\documentclass[trackchanges, twocolumn, shortbib]{aastex701}
\usepackage{amsmath, amssymb}
\usepackage{graphicx,color,float,tabularx,siunitx}
\definecolor{colorLink}{rgb}{0,0,180} 
\usepackage{hyperref}
\hypersetup{
   colorlinks = true,
   citecolor  = colorLink,
   urlcolor   = colorLink,
   linkcolor  = colorLink,
}

\begin{document}

\title{Study of Flat Spectrum Radio Quasars and BL Lacertae Objects as Sources of Diffusive Ultra High-Energy Cosmic Rays}

\author[orcid=0009-0003-2362-2080]{Swaraj~Pratim~Sarmah}
\affiliation{Department of Physics, Indian Institute of Technology, Guwahati, 781039, Assam, India}
\email[show]{swarajpratimsarmah16@gmail.com}  

\author[orcid=0000-0003-0012-7549]{Umananda~Dev~Goswami} 
\affiliation{Department of Physics, Dibrugarh University, Dibrugarh, 786004, Assam, India}
\email[show]{umananda@dibru.ac.in}

\begin{abstract}

We examine whether Flat Spectrum Radio Quasars (FSRQs) and BL Lacertae objects 
(BL Lacs) can act as plausible astrophysical sources of diffuse 
ultra-high-energy cosmic rays (UHECRs). Using realistic luminosity-dependent 
density evolution (LDDE) functions derived from observed gamma-ray luminosity 
functions for FSRQs and BL Lacs, we calculate the redshift evolution of the 
cosmic ray source population through integrated luminosity functions. The 
diffuse UHECRs flux from these sources is modelled by propagating nuclei 
through extragalactic space, including energy losses from interactions with 
cosmic photon backgrounds. The resulting UHECRs spectra are compared with 
observational data from the Pierre Auger Observatory and the Telescope Array, 
with fluxes normalised at reference energies. 
We also construct illustrative full skymaps of the integral UHECRs 
flux from $10$ EeV to $100$ EeV using the \textsc{HEALPix} framework, 
assuming a uniform distribution of synthetic sources and including
energy-dependent magnetic diffusion.
Our results show that although both FSRQs and BL Lac objects can reproduce 
the observed UHECR flux, the FSRQ scenario is disfavoured by anisotropy 
and inconsistent source density, while BL Lacs remain consistent with observations.

\end{abstract}

\keywords{{Ultra-High Energy Cosmic Rays}; {Flux} ; {Flat Spectrum Radio Quasars (FSRQs)} ; {BL Lacertae}}

\section{Introduction} \label{sec:intro}
The origin of ultra high-energy cosmic rays (UHECRs; $E\gtrsim 10^{17}$ eV) 
remains as one of the outstanding puzzles in modern astrophysics (for reviews, 
see \cite{Kotera:2011cp, Anchordoqui:2018qom}). Unlike photons and neutrinos, 
the arrival directions of UHECRs are obscured due to deflections in galactic 
and extragalactic magnetic fields \citep{sigl, globus_2008AA}, complicating 
the task of identifying their sources. Although the observed features in the 
UHECRs spectrum suggest an extragalactic origin, the precise energy at which 
the transition between galactic and extragalactic contributions occurs is 
still uncertain (e.g., see \cite{Aloisio:2012ba}). The Pierre Auger 
Observatory (Auger) in Argentina \citep{PierreAuger:2015eyc} and the Telescope 
Array (TA) in Utah, USA \citep{TelescopeArray:2015dcv}, currently provide the 
most detailed measurements of these particles.  

A few spectacular events recorded in different UHECR experiments already 
highlight the involvement of extreme energies in such events. E.g., the 
``Oh-My-God'' particle, detected by the Fly’s Eye experiment, carried an 
estimated energy of $\approx 320$~EeV ($1$ EeV $= 10^{18}$ eV), making it 
the most energetic event recorded via the 
fluorescence technique \citep{HIRES:1994ijd}. More recently, the TA 
Collaboration reported the ``Amaterasu'' event on May 27, 2021, which is the 
most energetic cosmic ray (CR) ever detected with a surface detector 
array. Its reconstructed energy of $244 \pm 29 \text{\ (stat)}^{+51}_{-76} 
\text{\ (sys)}$~EeV and arrival direction (RA, Dec) of 
($255.9^\circ\pm 0.6^\circ$, $16.1^\circ\pm 0.5^\circ$) places it among the 
most energetic particles observed to date \citep{TelescopeArray:2023sbd}.  
Such extreme-energy events raise the fundamental question about their 
acceleration mechanisms and astrophysical origins.  

Propagation effects, i.e., the effects of the environmental situations on CRs
while they propagate through space, also play a crucial role in their 
properties, including flux, energy and arrival directions from their sources. 
At the highest energies, UHECRs are expected to suffer significant attenuation 
due to the Greisen–Zatsepin–Kuz'min (GZK) suppression 
\citep{KGreisen1966, GZ1966}, 
arising from photopion production with the cosmic microwave background (CMB) 
photons. The attenuation length of protons at these energies is of order 
$\sim 10$~Mpc (e.g., see \cite{Dermer:2008cy}). Composition studies add 
further complexity: while TA results suggest predominantly light primaries 
at $E \gtrsim 10^{19}$ eV, significant uncertainties still remain 
\citep{HiRes:2009fiy}.  

Several astrophysical and cosmological scenarios have been proposed
to account for the sources of UHECRs, which include ultraheavy nuclei 
\citep{Zhang:2024sjp}, binary neutron star mergers \citep{Farrar:2024zsm}, 
the decay of superheavy dark matter \citep{Murase:2025uwv}, and transient
events in unresolved galaxies \citep{Unger:2023hnu}. 
In addition, cosmological scenarios based on modified or alternative gravity 
theories provide an alternative framework to explain the acceleration, 
propagation, and large-scale distribution of UHECRs in the evolving Universe 
\citep{Sarmah:2025uzk, Sarmah:2024kek}. Moreover, UHECRs are powerful probes 
of new physics such as Lorentz invariance violation (LIV) \citep{auger2022}, 
and this possibility has also been explored for the Amaterasu event 
\citep{Lang:2024jmc}. Recent high-precision measurements of UHECRs reveal 
several characteristic features: a hardening near the ankle at $\sim 5$ EeV
\citep{Bird:1993yi, hires2008, Abraham:2010mj}, a subsequent steepening at 
$\sim 13$ EeV \citep{hires2008, Auger2008}, and a strong suppression beyond 
$\sim 50$ EeV \citep{Auger2021,Verzi:2017hro}. Composition studies based on 
the depth of shower maximum, $X_{\rm max}$, show that UHECRs are relatively 
light at a few EeV and become progressively heavier with increasing energy. 
Moreover, the narrow dispersion in $X_{\rm max}$ above the ankle suggests 
that the mass distribution at a given energy is relatively tight.  

When these observations are interpreted within astrophysical source models, 
they indicate that two distinct source populations are required to explain the 
UHECRs flux above $\sim 1$~EeV and across the ankle \citep{Aloisio:2013hya, 
pao_jcap04, PierreAuger:2022atd}. The low-energy (LE) population, dominant 
below a few EeV, is likely composed of light and intermediate nuclei 
(H, He, N) with steep spectra $\propto E^{-\gamma_{\rm L}}$, where 
$\gamma_{\rm L}\simeq 3-3.7$, and with an only loosely constrained rigidity 
cutoff. At higher energies, the high-energy (HE) population becomes dominant, 
consisting of progressively heavier nuclei. This component is expected to have 
a much harder injection spectrum $\propto E^{-\gamma_{\rm H}}$, with 
$\gamma_{\rm H}<1$, and in some scenarios even negative spectral indices. 
Together, these populations provide a consistent description of the observed 
UHECRs spectrum and composition, linking spectral features to the physics of 
acceleration and propagation.  

Radio galaxies are promising candidates for the origin of UHECRs due 
to their powerful jets and large-scale structures capable of accelerating 
particles to ultra-high energies \citep{Seo:2025oiw}. Nearby sources such as 
Centaurus A, Virgo A, and Fornax A are particularly important, as their 
proximity allows them to contribute significantly to the observed flux and 
anisotropy \citep{Seo:2025oiw, Eichmann:2022ias}. In particular, Centaurus A 
has been suggested as a possible contributor to the observed dipole 
anisotropy \citep{Wang:2024ijr}. Combined studies of spectrum, composition, 
and anisotropy indicate that a few nearby radio galaxies may dominate the 
UHECR flux at the highest energies, while more distant sources contribute to 
an approximately isotropic background \citep{Eichmann:2022ias}. Recent results 
from the Auger further support this picture, favoring contributions from 
nearby sources such as starburst galaxies and Centaurus A, while disfavoring 
models based on $\gamma$-ray selected AGN populations 
\citep{PierreAuger:2023htc}. Nevertheless, AGN, including blazars, remain 
important to study, as they are powerful particle accelerators and may still 
contribute to the diffuse UHECR flux, particularly within propagation-based 
models and in the context of multi-messenger observations involving 
high-energy neutrinos and gamma rays 
\citep{Das:2020nvx, Murase:2011cy, Das:2025tfq}.

The study of radio-loud (RL) active galactic nuclei (AGN), particularly 
blazars with strongly relativistic and beamed jets, provides a unique method 
to probe jet activity, black hole spin, and the role of major mergers in 
galaxy evolution. A key statistical tool in this context is the luminosity 
function (LF) of blazars, defined as the number of blazars per comoving volume 
within a given luminosity interval, and its evolution with redshift.  
The Fermi Gamma-ray Space Telescope ({\it Fermi-LAT}) has revolutionized this 
field: thanks to its sensitivity and all-sky uniform coverage, it has detected 
hundreds of blazars spanning from the local Universe up to 
$z = 3.1$ \citep{Fermi-LAT:2011lqa}. The blazar LF not only constrains their 
contribution to the diffuse extragalactic $\gamma$-ray background but also 
helps establish connections between blazars and their parent AGN populations 
\citep{Ajello:2008xb, Inoue:2008pk}. Extensive multiwavelength studies of 
blazars have been carried out in the radio \citep{Dunlop90, Wall:2004tg} and 
soft X-ray~\citep{giommi94, Rector:2000xr, Wolter:2001ri, Caccianiga:2001vu, 
Beckmann:2003fy, Padovani:2007qb} bands. Flat-spectrum radio quasars (FSRQs), 
the most luminous subclass of blazars, exhibit strong positive cosmological 
evolution, meaning they were more numerous in the past \citep{Dunlop90}. 
This trend persists up to a luminosity-dependent redshift cutoff 
(e.g., see \cite{Padovani:2007qb, Ajello:2008rd}). Interestingly, FSRQs 
follow an evolutionary pattern similar to that of X-ray-selected, radio-quiet 
AGNs \citep{Ueda:2003yx, Hasinger:2005sb, LaFranca:2005qs}, suggesting 
possible common mechanisms driving their cosmic evolution.  

Another important sub-population of blazars are BL Lacertae (BL Lac) objects 
(e.g., see \citep{blandford78}). They represent an extreme class of AGNs 
characterized by highly variable emission, most likely produced by 
relativistic jets aligned very close to our line of sight. BL Lac objects 
are distinguished from their FSRQ counterparts by their optical spectra, 
which lack significant emission lines with equivalent width $>5 \AA$ 
(e.g., see \cite{Urry:1995mg, marcha96}). Their spectra are typically 
dominated by a power-law continuum, indicating either exceptionally strong 
non-thermal jet emission (aligned nearly directly with our line of sight), 
or unusually weak thermal disk and broad-line emission, the latter plausibly 
linked to low accretion activity \citep{Giommi:2013mck}. 

Considering the said interesting nature of emission with high redshift 
values of FSRQs and BL Lac objects, we aim to investigate these objects 
as plausible sources of UHECRs from the perspective of the characteristics of 
UHE fluxes from them. Thus, the remainder of the paper is organised as follows. 
Section~\ref{source} is 
divided into two subsections. In Subsection~\ref{fsrq}, we describe the 
evolution of FSRQ sources, while Subsection~\ref{bllac} focuses on the 
evolution of BL Lac objects. Section~\ref{secIII} presents the basic equations 
governing the propagation of UHECRs. In Section~\ref{flux_results}, we provide 
the flux calculation equations, discuss the corresponding mixed composition 
results, generate skymaps, identify hotspots, and locate nearby galaxies 
within $1^\circ$ angular separation. Finally, Section~\ref{summary} summarises 
the findings and draws conclusions of the study.

\section{Evolution of AGN Sources}\label{source}
In this section, we describe the evolution of two classes of AGN sources, 
viz., the FSRQs and BL Lac objects, which we consider as potential sources of 
UHECRs, as mentioned already. First, we describe the FSRQs' evolution, and in 
the next the evolution of the BL Lac objects is discussed as follows.
\subsection{Flat Spectrum Radio Quasars (FSRQs)}\label{fsrq}
The space density of radio-quiet AGNs is observed to reach a maximum at 
intermediate redshifts. The epoch of this `redshift peak' is correlated with 
the luminosity of the sources (e.g., see \cite{Ueda:2003yx, Hasinger:2005sb}).
This peak likely reflects the combined effects of supermassive black hole 
(SMBH) growth over cosmic time and a decline in fuelling activity as the 
frequency of major mergers decreases at later times. To investigate whether 
similar behaviour occurs in the {\it Fermi-LAT} FSRQ population, Ajello et 
al.~fit the data using a pure luminosity evolution (PLE) model as described in 
Ref.~\cite{Ajello:2011zi} and is given as
\begin{equation}
\Phi(L_{\gamma},z) = \Phi(L_{\gamma}/e(z)),
\label{eq:ple}
\end{equation}
where\\[-15pt]
\begin{align}
\Phi(L_{\gamma}/e(z=0)) = \frac{dN}{dL_{\gamma}}=\;&\frac{A}{2.303\,L_{\gamma}} \nonumber \\
&\times \left[\left(\frac{L_{\gamma}}{L_{*}}\right)^{\gamma_1}+
\left(\frac{L_{\gamma}}{L_{*}}\right)^{\gamma2} 
\right]^{-1}
\label{eq:ple1}
\end{align}
\vspace{-5pt}
and 
\begin{equation}
e(z) = (1+z)^k e^{z/\xi}.
\label{eq:ple2}
\end{equation}
In the PLE model, the comoving luminosity function 
$\Phi(L_{\gamma},z)$ is assumed to evolve only in luminosity, such that the 
redshift dependence enters through the evolution factor $e(z)$. Here, 
$L_{\gamma}$ denotes the $\gamma$-ray luminosity of the source and 
$\Phi(L_{\gamma},z)$ gives the number density of sources per unit luminosity.
It is to be noted that the observed $\gamma$-ray luminosities of 
blazars are affected by Doppler boosting and hence $L_{\gamma}$ is related 
with the intrinsic luminosity $L_{\gamma}'$ as 
$L_{\gamma} = (\delta_{\rm e}^{6}/\Gamma_{\rm e}^{2}) L_{\gamma}'$ 
\citep{Ajello:2011zi, Das:2020hev}, where $\delta_{\rm e}$ is the Doppler 
factor of the jet, which quantifies the enhancement of the observed luminosity 
$L_{\gamma}$ due to relativistic motion, and $\Gamma_{\rm e}$ is the Lorentz 
factor.
The local luminosity function at $z=0$ is characterized by a broken power-law 
with a normalization factor $A$, break luminosity $L_{*}$, low-luminosity 
slope $\gamma_{1}$, and high-luminosity slope $\gamma_{2}$, where $\gamma_{1}$ 
controls the behavior of the faint-end population ($L_{\gamma}\ll L_{*}$) and 
$\gamma_{2}$ determines the fall-off at the bright end ($L_{\gamma}\gg L_{*}$).
The luminosity evolution factor, $e(z)=(1+z)^{k}\exp(z/\xi)$, describes how 
the characteristic luminosity shifts with redshift, where $k$ sets the 
strength of luminosity evolution and $\xi$ controls the redshift scale at 
which the evolution transits or saturates. As a simple PLE function model does 
not adequately reproduce the 
{\it Fermi-LAT} data, and given evidence for a luminosity-dependent evolution 
of the redshift peak, we fit the {\it Fermi-LAT} FSRQ sample using a 
luminosity-dependent density evolution (LDDE) model. In this approach, the 
evolution occurs mainly in density, with the redshift peak depending on 
luminosity. The LDDE model is parametrised as in Ref.~\cite{Ajello:2011zi},
which is
\begin{equation}
\Phi(L_{\gamma},z) = \Phi(L_{\gamma}) \times e(z,L_{\gamma}),
\label{ldde}
\end{equation}
where\\[-15pt]
\begin{equation}
e(z,L_{\gamma})= \left[ 
\left( \frac{1+z}{1+z_c(L_{\gamma})}\right)^{p1} + 
\left( \frac{1+z}{1+z_c(L_{\gamma})}\right)^{p2} 	
 \right]^{-1}
\label{eq:evol}
\end{equation}
\vspace{-5pt}
and 
\begin{equation}
z_c(L_{\gamma})= z_c^*\cdot (L_{\gamma}/10^{48})^{\alpha}.
\label{eq:zpeak}
\end{equation}
Here, $z_c(L_{\gamma})$ represents the luminosity-dependent redshift at which 
the evolutionary trend changes sign (from positive to negative), with $z_c^*$ 
denoting the redshift peak for an FSRQ of luminosity $10^{48}$\,erg s$^{-1}$ 
and the parameter $\alpha$ controls how the peak redshift shifts with 
luminosity. The evolution factor $e(z,L_{\gamma})$ is parametrized by two 
evolutionary indices, $p_{1}$ and $p_{2}$, which describe the redshift 
evolution below and above the luminosity-dependent peak redshift 
$z_{c}(L_{\gamma})$, respectively. 
The values of all parameters are adopted from Ref.~\cite{Ajello:2011zi}.
$\Phi(L_{\gamma})$ adopts the same double power-law form as in 
Eq.~\eqref{eq:ple1}. This parametrisation is similar to that proposed by 
Ref.~\cite{Ueda:2003yx}, but it is continuous around the redshift peak 
$z_c(L_{\gamma})$. Such continuity is advantageous for fitting algorithms that 
use derivatives of the function to locate the minimum.

\subsection{BL Lacertae (BL Lac) Objects}\label{bllac}
In this case, we use the intrinsic distribution of photon indices following 
Ref.~\cite{Ajello:2013lka}, where the photon index $\Gamma$ 
is assumed to follow a Gaussian distribution for a given redshift 
$z$ and luminosity $L_{\gamma}$ and hence the luminosity function becomes
\begin{equation}
\Phi(L_{\gamma},z,\Gamma) \propto \exp\left[-\frac{(\Gamma - \mu(L_{\gamma}))^{2}}{2\sigma^{2}}\right],
\label{eq:index}
\end{equation}
where $\mu$ and $\sigma$ are the Gaussian mean and 
dispersion, respectively. Previous works \citep{Ghisellini:2009rb, Meyer:2012jf}
have suggested that the mean photon index may correlate with 
luminosity, and thus we allow $\mu$ to vary with $L_{\gamma}$ 
following the blazar sequence prescription of Ref.~\cite{Ajello:2013lka}, 
given by
\begin{equation}
\mu(L_{\gamma}) = \mu^{*} + \beta\,[\log_{10}(L_{\gamma}) - 46],
\label{eq:blazseq}
\end{equation}
where $\mu^{*}$ is the mean photon index at a reference 
luminosity $L_{\gamma} = 10^{46}\,\mathrm{erg\,s^{-1}}$, and 
$\beta$ quantifies the slope of the luminosity–spectral index correlation.
The LF at redshift $z=0$ is modelled as a smoothly connected double power law, 
multiplied by the photon index distribution, given in Eq.~\eqref{eq:index}, as
\begin{align}
\Phi(L_{\gamma},z=0, \Gamma) =\frac{A}{2.303\,L_{\gamma}}&
\left[\left(\frac{L_{\gamma}}{L_{*}}\right)^{\gamma_1}+
\left(\frac{L_{\gamma}}{L_{*}}\right)^{\gamma2} 
\right]^{-1} \nonumber \\[8pt]  
&\times \exp\left[-\,\frac{ (\Gamma-\mu(L_{\gamma}))^2}{2\sigma^2}\right].
\label{eq:lf0}
\end{align}
As in the FSRQs case, in this case the relation between the observed 
luminosity $L_{\gamma}$ and the intrinsic luminosity $L_{\gamma}'$ is
$L_{\gamma} = \delta_{\rm e}^{4} L_{\gamma}'$ 
\citep{Ajello:2013lka,Das:2020hev}. Also, we utilise here the same formalism 
as performed in Eqs.~\eqref{ldde}--\eqref{eq:zpeak} and the best fit
values of the parameters are adopted from Ref.~\cite{Ajello:2013lka}.

\section{Diffusion of Cosmic Rays in Turbulent Magnetic Fields}\label{secIII}
Modelling extragalactic magnetic fields remains a complex task due to several 
uncertainties and limitations \citep{han}. The exact strength and structure 
of these fields are still not well known and can vary greatly across different 
parts of extragalactic space \citep{hu_apj, urmilla}. To make the study more 
manageable, we consider the propagation of CRs in a uniform and turbulent 
extragalactic magnetic field. Such a field is described by two key parameters: 
its root mean square (RMS) strength $B$ and its coherence length $l_\text{c}$. 
The RMS strength, defined as $\sqrt{\langle B^2(x)\rangle}$, may be as weak as 
$10^{-16}$~G in cosmic voids \citep{Neronov:2010gir}, increase up to 
$\gtrsim 10$~nG in galaxy clusters, and typically reach about $1$~nG in 
filaments \citep{feretti, Valle, Vazza}. The coherence length $l_\text{c}$ 
generally lies between $0.01$~Mpc and $1$~Mpc~\citep{sigl}. The effective 
Larmor radius $r_\text{L}$ of a charged particle with charge $Ze$ and energy 
$E$ moving through a turbulent magnetic field of strength $B$ can then be 
written as
\begin{equation}\label{larmor} r_\text{L} = \frac{E}{ZeB} \simeq 1.1\, \frac{E/\text{EeV}}{ZB/\text{nG}}\;\text{Mpc}. \end{equation}

The idea of critical energy is essential for describing how charged particles 
diffuse in magnetic fields. It is the energy at which the coherence length of 
a particle with charge $Ze$ becomes equal to its Larmor radius, i.e., 
$r_\text{L}(E_\text{c}) = l_\text{c}$. Hence, the critical energy is written 
as
\begin{equation}\label{cri_energy}
E_\text{c} = ZeBl_\text{c} \simeq 0.9 Z\, \frac{B}{\text{nG}}\, \frac{l_\text{c}}{\text{Mpc}}\;\text{EeV}.
\end{equation}
This energy separates two different diffusion regimes: resonant diffusion at 
energies below $E_\text{c}$ and non-resonant diffusion at energies above 
$E_\text{c}$.

The energy-dependent diffusion coefficient $D$ is expressed as \citep{harari}
\begin{equation}\label{diff_coeff}
D(E) \simeq \frac{c\,l_\text{c}}{3}\left[4 \left(\frac{E}{E_\text{c}} \right)^2 + a_\text{I} \left(\frac{E}{E_\text{c}} \right) + a_\text{L} \left(\frac{E}{E_\text{c}} \right)^{2-\gamma} \right],
\end{equation}
where $\gamma$ is the spectral index, and $a_\text{I}$ and $a_\text{L}$ are 
constants. For a Kolmogorov-type turbulence spectrum, $\gamma = 5/3$ with $a_\text{L} \approx 0.23$ and $a_\text{I} \approx 0.9$. As noted earlier, the 
diffusion length $l_\text{D}$ represents the typical distance over which a 
particle’s total deflection becomes roughly one radian, defined as 
$l_\text{D} = 3D/c$. Throughout this work, we assume a homogeneous 
turbulent intergalactic magnetic field with RMS strength 
$B_{\rm rms}=1~\mathrm{nG}$ and coherence length $l_{\rm c}=0.5~\mathrm{Mpc}$, 
corresponding to the maximum scale of a Kolmogorov turbulence spectrum.

In the diffusive regime, the transport equation for UHE particles moving 
through an expanding Universe from a source at position $x_\text{s}$ is 
given by \citep{berezinskyGre}
\begin{align}\label{diff_eqn}
\frac{\partial n}{\partial t} + 3 H(t)\, n - b(E,t)\, \frac{\partial n}{\partial E}& -n\, \frac{\partial n}{\partial E} - \frac{D(E,t)}{a^2(t)}\,\nabla^2 n  \nonumber \\
&= \frac{\mathcal{N}(E,t)}{a^3(t)}\,\delta^3({x}-{x}_\text{s}),
\end{align}
where the Hubble parameter is $H(t) = \dot{a}(t)/a(t)$, with $\dot{a}(t)$ 
representing the rate of change of the scale factor $a(t)$ over cosmic time 
$t$. The coordinates ${x}$ denote comoving positions, $n$ is the particle 
density, and $\mathcal{N}(E)$ represents the source emissivity. At a time $t$ 
corresponding to redshift $z$, the separation between the source and the 
particle is given by $r_\text{s} = {x} - {x_\text{s}}$.
The propagation of UHECRs is governed by continuous energy losses arising from cosmological expansion and interactions with ambient photon backgrounds. The general energy loss equation can be written as
\begin{equation}
\frac{dE}{dt} = -\, b(E,t),
\qquad
b(E,t) = H(t)E + b_{\rm int}(E,t),
\label{eq:energy_loss_general}
\end{equation}
where the term $H(t)E$ describes adiabatic energy losses due to the expansion of the Universe, while $b_{\rm int}(E,t)$ accounts for interaction losses with background radiation fields.
The interaction term is redshift dependent and includes contributions from the cosmic microwave background (CMB) and the extragalactic background light (EBL). It can be written as \citep{Aloisio:2008pp}
\begin{align}
b_{\rm int}(E,z) = b_{\rm pair}(E,z) \, + \, b_{\pi}(E,z)\, +&  \, b_{\rm dis}^{\rm CMB}(E,z)\,  \nonumber \\
&+ \,b_{\rm dis}^{\rm EBL}(E,z),
\label{eq:bint_total}
\end{align}
where $b_{\rm pair}$ denotes Bethe-Heitler pair production, $b_{\pi}$ corresponds to photopion production (relevant for protons), and $b_{\rm dis}^{\rm CMB}$ and $b_{\rm dis}^{\rm EBL}$ represent photodisintegration losses of nuclei due to interactions with CMB and EBL photons, respectively \citep{Aloisio:2008pp}. For proton primaries ($A=1$), the dominant processes are pair production and photopion production on the CMB, treated within the continuous energy–loss approximation. For nuclei ($A>1$), photodisintegration becomes important in addition to pair production; this process changes the mass number while approximately conserving the Lorentz factor.
The general photodisintegration rate for a nucleus of Lorentz factor $\Gamma = E/(A m_p c^2)$ interacting with an isotropic photon background of spectral density $n(\varepsilon,z)$ is given by \citep{Aloisio:2010he}
\begin{equation}
\frac{1}{\tau_{A\gamma}(\Gamma,z)}
=
\frac{c}{2\Gamma^2}
\int_{\varepsilon_{\rm th}}^{\infty}
d\varepsilon_r \,
\sigma_{A\gamma}(\varepsilon_r)
\, \varepsilon_r
\int_{\varepsilon_r/(2\Gamma)}^{\infty}
d\varepsilon \,
\frac{n(\varepsilon,z)}{\varepsilon^{2}},
\label{eq:photodisintegration_rate}
\end{equation}
where $\varepsilon_r$ is the photon energy in the nuclear rest frame, $\varepsilon$ is the photon energy in the laboratory frame, $\varepsilon_{\rm th}$ is the photodisintegration threshold, and $\sigma_{A\gamma}(\varepsilon_r)$ is the corresponding cross section. The dominant contribution arises from the Giant Dipole Resonance, which can be modelled with a Lorentzian profile centered at $\varepsilon_0$.
Since photodisintegration alters the mass number $A$ but leaves the Lorentz factor approximately unchanged, the associated continuous energy-loss term can be written as \citep{Aloisio:2010he}
\begin{equation}
b_{\rm dis}(E,z)
=
\frac{E}{A}
\sum_i i \, R_i(\Gamma,z),
\label{eq:energy_loss_disintegration}
\end{equation}
where $R_i(\Gamma,z)$ denotes the partial interaction rate for emission of $i$ nucleons.
For the CMB, the photon distribution follows a Planck spectrum with temperature evolving as $T(z)=T_0(1+z)$. For the EBL, one adopts evolutionary models in which the photon density evolves according to \citep{Aloisio:2010he}
\begin{equation}
n_{\rm EBL}(\varepsilon,z)
=
(1+z)^{-3/2}\, n_0(\varepsilon),
\label{eq:ebl_evolution}
\end{equation}
where $n_0(\varepsilon)$ is the locally observed EBL spectrum.
The EBL contribution to the photodisintegration energy loss entering Eq.~(\ref{eq:bint_total}) is obtained by substituting $n(\varepsilon,z)=n_{\rm EBL}(\varepsilon,z)$ into Eq.~(\ref{eq:photodisintegration_rate}).
For numerical implementation, it is convenient to express the evolution equation in terms of redshift. Following the Ref.~\citep{harari, swaraj1, molerach}
\begin{equation}
\frac{dt}{dz} = -\frac{1}{(1+z)H(z)},
\end{equation}
the energy evolution becomes \citep{harari}
\begin{equation}
\frac{dE}{dz}
=
\frac{E}{1+z}
+
\frac{b_{\rm int}(E,z)}{(1+z)H(z)}.
\label{eq:dEdz}
\end{equation}
Equation~(\ref{eq:dEdz}) is integrated from the observer at $z=0$ to a source at redshift $z_s$ to determine the injection energy required for a particle observed today with energy $E$. The generation energy $E_g$ at redshift $z$
to the observed energy $E$ at $z=0$ is obtained by differentiating the
energy-evolution equation with respect to $E$ \citep{harari}. The result is
\begin{equation}
\frac{dE_g}{dE}
=
(1+z)
\exp\!\left[
\int_{0}^{z}
\left|
\frac{dt}{dz'}
\right|
\,
\frac{\partial b_{\rm int}(E(z'),z')}{\partial E}
\, dz'
\right],
\label{eq:jacobian_general}
\end{equation}
where $E(z')$ is the solution of Eq.~(\ref{eq:dEdz}) along the
characteristic trajectory connecting $E$ at $z=0$ to $E_g$ at redshift $z$.

This treatment consistently accounts for cosmological expansion, CMB and
EBL photon backgrounds, and the dominant interaction processes governing
UHECR propagation.
This procedure consistently accounts for adiabatic losses and interaction
losses during propagation. According to the 
Ref.~\citep{mollerachjcap}, photodisintegration largely preserves the
Lorentz factor and rigidity of the main nuclear fragment. As a result,
this process does not change the diffusion coefficient or the overall
form of particle propagation, and its effect enters only through the
nuclear survival probability. Similar treatments have also been discussed in 
the context of modified and alternative gravity models
\citep{swaraj2, swaraj3, Sarmah:2024kek}.

The general solution to Eq.~\eqref{diff_eqn} was derived in 
Ref.~\cite{berezinskyGre} and is written as
\begin{equation}\label{density}
n(E,r_\text{s})= \int_{0}^{z_{i}} dz\, \bigg | \frac{dt}{dz} \bigg |\, \mathcal{N}(E_\text{g},z)\, \frac{\textrm{exp}\left[-r_\text{s}^2/4 \lambda^2\right]}{(4\pi \lambda^2)^{3/2}}\, \frac{dE_\text{g}}{dE},
\end{equation}
where $\lambda$ is the Syrovatskii variable, defined as 
\citep{syrovatsky_1959}
\begin{equation}\label{syro}
\lambda^2(E, z) = \int_{0}^{z} dz \left| \frac{dt}{dz} \right| (1 + z)^2 D(E_\text{g}, z).
\end{equation}
Here, $E_\text{g}(E, z)$ is the generation energy at redshift $z$ 
corresponding to an observed energy $E$ at $z = 0$. The relation
between cosmological time and redshift for standard $\Lambda$CDM
model is given by
\begin{equation}
\left| \frac{dt}{dz}\right|=\frac{1}{H_0\,(1+z)\,\sqrt{(1+z)^3\,\Omega_m+ \Omega_\Lambda}},
\end{equation}
where, $H_0=67.4$\,km s$^{-1}\,$Mpc$^{-1}$, $\Omega_m=0.315$, and $\Omega_\Lambda=0.699$ according to the Planck's data \citep{planck2018}.
In the diffusive regime, 
the particle density depends on the particle energy, the distance from the 
source, and the properties of the turbulent magnetic field (TMF). Diffusion 
can cause the observed density of CRs at a given distance to be considerably 
higher than what would be expected for rectilinear propagation, which follows 
a $1/r^2$ law. This enhancement results from delayed escape and multiple 
scattering of particles, and is quantified by the enhancement factor, defined 
as the ratio of the diffusive density to the rectilinear density 
\citep{molerach} as given by
\begin{equation}\label{enhancement}
\xi(E,r_\text{s})=\frac{4\pi r_\text{s}^2c\, n(E, r_\text{s})}{\mathcal{N}(E)}.
\end{equation}

\section{Flux of UHECRs}\label{flux_results}
The diffusion of CRs in TMFs has been studied extensively by many authors 
\citep{berezinskyGre, blasi, Sarmah:2025yoy, globus, sigl, stanev, kotera, 
swaraj5, Yoshiguchi, lemoine1, hooper, hooper2, sigl2007, aloisio_ptep, 
swaraj1}. Berezinsky and Gazizov \citep{berezinsky_gazizov, berezinskyGre} 
generalised the Syrovatskii solution \citep{syrovatsky_1959} to examine the 
diffusion of protons in an expanding Universe. The flux from a CR source at a 
distance $r_\text{s}$, much larger than the diffusion length $l_\text{D}$, 
can be obtained by solving the diffusion equation in an expanding Universe 
\citep{berezinskyGre}, leading to the expression \citep{Manuel}:
\begin{align}\label{fluxeq}
J(E) = \frac{c}{4\pi} \int_{0}^{z_{\text{max}}}  dz \,&  \left| \frac{dt}{dz} \right| \, \mathcal{N}\left[E_\text{g}(E, z), z\right]\\ \nonumber
&\frac{\exp\left[-r_\text{s}^2 / (4 \lambda^2)\right]}{(4 \pi \lambda^2)^{3/2}} \frac{dE_\text{g}}{dE},
\end{align}
where $z_{\text{max}}$ is the maximum redshift at which the source starts 
emitting CRs, and $E_\text{g}$ is the generation energy at redshift $z$ 
corresponding to energy $E$ at $z = 0$.
The total source emissivity $\mathcal{N}$ is obtained by summing the 
charge-specific emissivities $\mathcal{N}_\text{Z}$ for different nuclei. The 
charge-specific emissivity follows a power-law with a rigidity dependent
cutoff at the maximum energy $E_\text{max}=Z E_p$, where $E_p$ is the maximum proton
energy, and is given by \citep{mollerachjcap}
\begin{equation}
\mathcal{N}_\text{Z}(E, z) = \varepsilon_\text{Z}
f(z) E^{-\gamma} / \cosh(E / E_{max}),
\end{equation}
where $\varepsilon_\text{Z}$ denotes the relative fractional 
contributions of nuclei with charge $Z$ to the CR flux and these fractions 
are treated as free parameters (constrained such that 
$\sum \varepsilon_\text{Z} = 1$). In this work, we adopt an effective 
parametrisation in which the maximum energy is expressed as 
\citep{Rieger:2022qhs}
\begin{equation}
E_{\max}(Z, L) = \eta \, Z \, E_0 \left(\frac{L}{L_0}\right)^{1/2},
\end{equation}
where $L$ is the source luminosity and $\eta < 1$ denotes the acceleration efficiency. We take 
$E_0 = 10^{20}~\mathrm{eV}$ and $L_0 = 10^{45}~\mathrm{erg\,s^{-1}}$, 
consistent with typical astrophysical accelerators \citep{Rieger:2022qhs}.
Furthermore, $f(z)$ describes the evolution of source emissivity with redshift 
$z$. In our study, we improve the description of $f(z)$ by adopting the LDDE 
model for FSRQs and BL Lacs derived by Ajello et 
al.~\citep{Ajello:2011zi, Ajello:2013lka}. Instead of using a simple 
phenomenological form, often taken as a power-law or broken power-law 
in $(1+z)$, we employ a physically motivated redshift evolution based on the 
observed gamma-ray luminosity function. In this approach, the evolution factor is expressed as
\begin{equation}
f(z) = \int_{L_{\min}}^{L_{\max}} \Phi(L, z)\, L\, dL,
\end{equation}
where $\Phi(L, z)$ is the luminosity function of the sources and 
$L_{\min}$, $L_{\max}$ are the luminosity limits. This formulation incorporates both the redshift and luminosity dependence of the AGN population in a self-consistent manner.

In our framework, the normalization of the UHECR flux is related to the gamma-ray luminosity function through a proportionality parameter, interpreted as the baryon loading factor $\kappa$, defined as
\begin{equation}
\kappa = \frac{\mathcal{L}_{\rm CR}}{\mathcal{L}_{\gamma}},
\end{equation}
where $\mathcal{L}_{\gamma}$ and $\mathcal{L}_{\rm CR}$ denote the local (i.e., $z = 0$) gamma-ray and CR luminosity densities, respectively. While the luminosity function determines the redshift evolution of the source emissivity, the overall normalization of the UHECR flux is governed by $\kappa$, which is constrained by fitting the predicted flux through the model parameters (source separation distance $d_\text{s}$, nuclei fraction $\varepsilon_Z$, and $E_\text{max}$ ) to the Auger and Telescope Array data.
This approach ensures that the source evolution is not arbitrary, but directly informed by observational data, providing a more realistic representation of UHECR source emissivity. In contrast, simple phenomenological forms such as $f(z) \propto (1+z)^m$ or broken power-laws do not account for the luminosity dependence of AGNs and may lead to oversimplified or biased estimates of source evolution.

Equation~\eqref{fluxeq} can be generalised to nuclei in terms of their 
rigidity \citep{mollerachjcap}.
For multiple sources, we apply the propagation theorem \citep{aloisio} to sum 
contributions from all sources, which satisfies
\begin{equation}\label{lim1}
\int_{0}^{\infty} dr \, 4\pi r^2 \frac{\exp\left[-r^2 / (4 \lambda^2)\right]}{(4 \pi \lambda^2)^{3/2}} = 1.
\end{equation}

To examine the effect of finite source distances on flux suppression, we compute the sum over a given set of distance distributions. These assume a uniform source density, with distances from the observer given by \citep{mollerachjcap, Manuel, swaraj7}
\begin{equation}
r_\text{i} = \left(\frac{3}{4\pi}\right)^{1/3}\, d_\text{s}\, \frac{\Gamma(i + 1/3)}{(i - 1)!},
\end{equation}
where $d_\text{s}$ is the separation distance between sources. For a discrete 
source distribution, summing over all sources yields a suppression factor 
\citep{mollerachjcap, Manuel}
\begin{equation} \label{F_supp}
F \equiv \frac{1}{n_\text{s}} \sum_i \frac{\exp\left[-r_\text{i}^2/4 \lambda^2\right]}{(4\pi \lambda^2)^{3/2}},
\end{equation}
in place of Eq.~\eqref{lim1}, where $n_\text{s}$ is the source density. Using 
Eq.~\eqref{fluxeq}, after summing over all sources, the modified flux for an 
ensemble of sources can then be expressed as
\begin{align}\label{flux}
J_\text{mod}(E)  \simeq  \frac{R_\text{H} n_\text{s}}{4\pi}\!& \int_{0}^{z_\text{max}}\!\!\!\!\! dz\, (1+z)^{-1}  \Big | \frac{dt}{dz} \Big | \nonumber \\
&\times \mathcal{N}\left[E_\text{g}(E, z), z\right]\, \frac{dE_\text{g}}{dE}\, F,
\end{align}
where $R_\text{H} = c/H_0=4.3$ Gpc is known as the Hubble radius and $z_\text{max}=0.5$. We can 
rewrite Eq.~\eqref{syro} in terms of $R_\text{H}$ and from 
Eq.~\eqref{diff_coeff} as
\begin{align}\label{ad}
\lambda^2(E,z) = &\,\frac{H_0R_\text{H} l_\text{c}}{3} \int_{0}^{z} \!\! dz\, \bigg| \frac{dt}{dz} \bigg| (1+z)^2 
\bigg[4 \left( \frac{(1+z)\,E}{E_\text{c}} \right)^2  \nonumber \\
    & + a_\text{I} \left( \frac{(1+z)\,E}{E_\text{c}} \right) 
    + a_\text{L} \left( \frac{(1+z)\,E}{E_\text{c}} \right)^{2-\gamma}
\bigg].
\end{align}
\begin{figure*}
\centerline{
\includegraphics[scale=0.3]{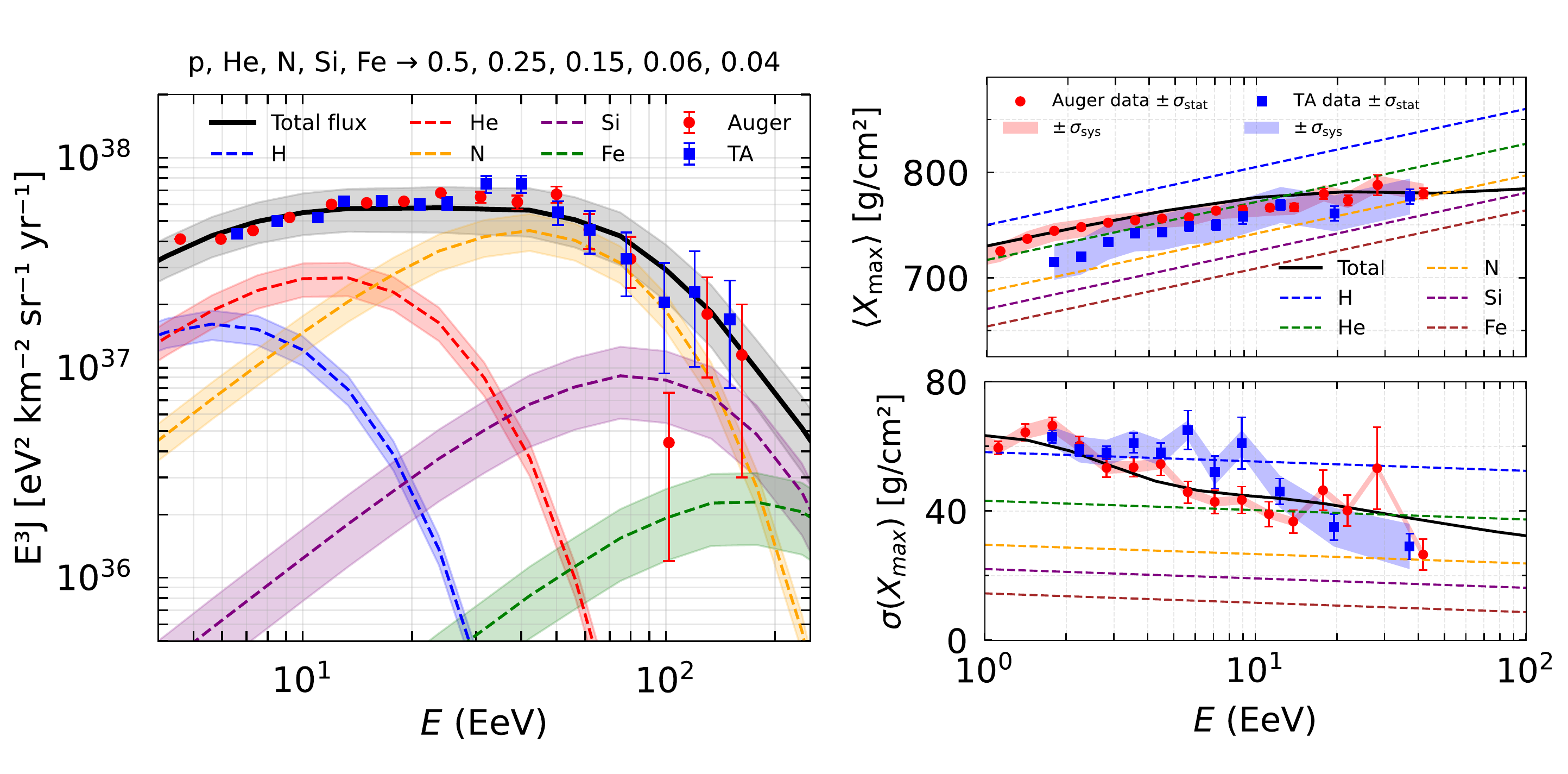}}
\centerline{
\includegraphics[scale=0.3]{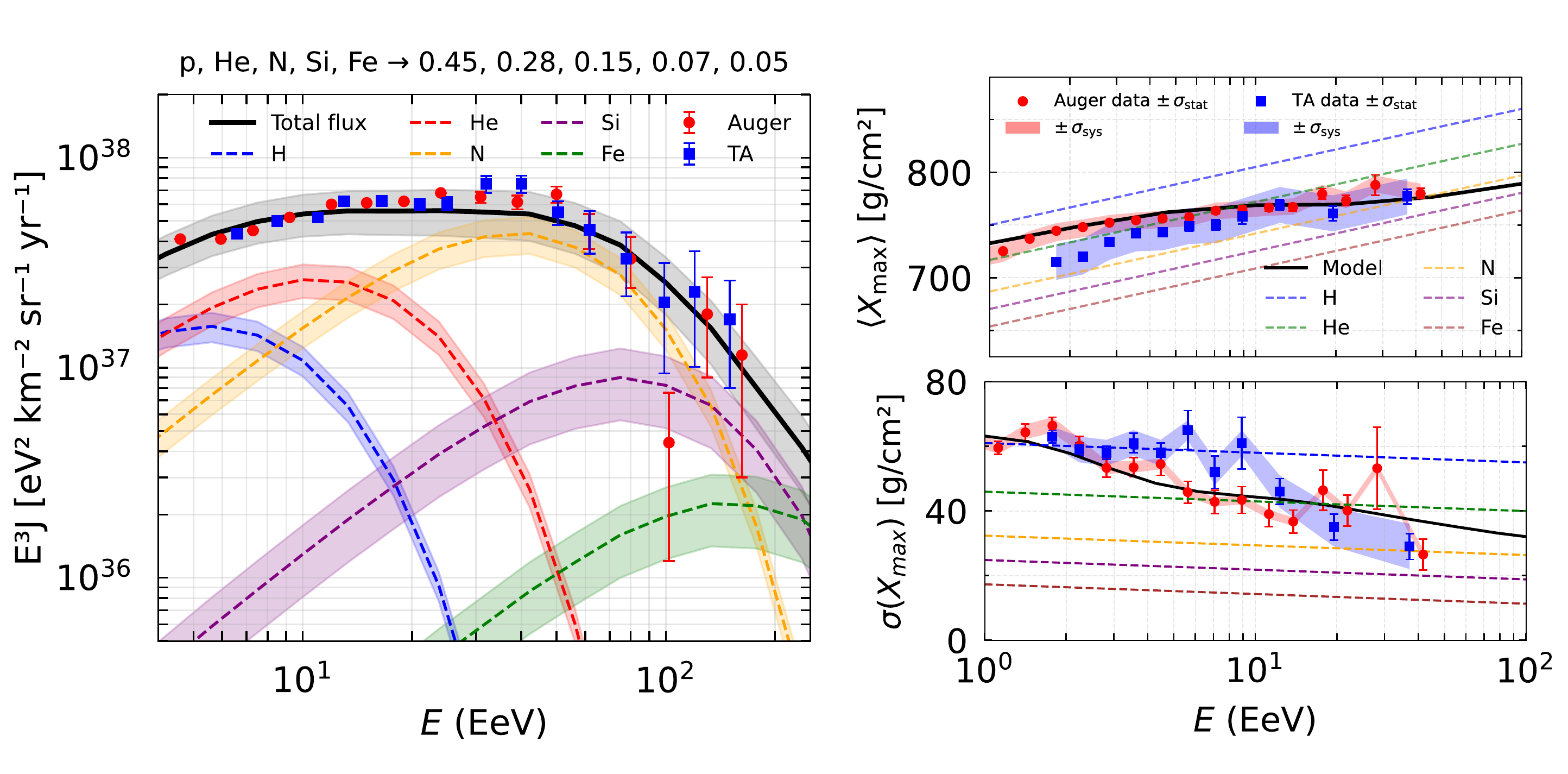}}
\caption{Fluxes of UHECRs and $X_\text{max}$ values for a mixed composition 
scenario from FSRQs (top panel) and BL Lac objects (bottom panel), obtained
by using different source separation distances (see text) and varying relative 
abundances of nuclear species. The observational values of flux of these two 
experiments are taken from Ref.~\cite{bergmanepj} and those for $X_\text{max}$ 
are from Ref.~\cite{xmax_pao, xmax_ta}.}
\label{mixed1}
\end{figure*}
For the mixed composition case, we draw Fig.~\ref{mixed1} for the fluxes 
along with the mean depths of shower maximum $\langle X_\text{max}\rangle$ as 
predicted by our theoretical calculations. The nuclear abundance of p, He, N, 
Si and Fe is mentioned at the top of the plots. Since we use two different AGN 
source population, so we get two different separation distance in fitting the 
numerical results with
Auger and TA data \citep{bergmanepj}. The depth of the shower maximum, $X_\text{max}$, is 
determined using a parametrization based on air shower physics 
\citep{Auger2010, gaisser1990}. For a nucleus with mass number $A$ and 
energy $E$, it is expressed as \citep{Auger2010, gaisser1990}.
\begin{equation}
X_\text{max}(E,A) = X_0 + \nu~ \text{log}_{10}\left(\frac{E}{A}\right),
\end{equation}
where $X_0$ and $\nu$ are parameters influenced by hadronic interactions
\citep{Auger2010}. In our analysis, we have used the values as
$X_0 = 700\, \text{g cm}^{-2}$ and $\nu = 50\, 
\text{g cm}^{-2}$ \citep{gaisser1990, gaisser2016}.
\begin{figure*}[!ht]
\centerline{
\includegraphics[scale=0.35]{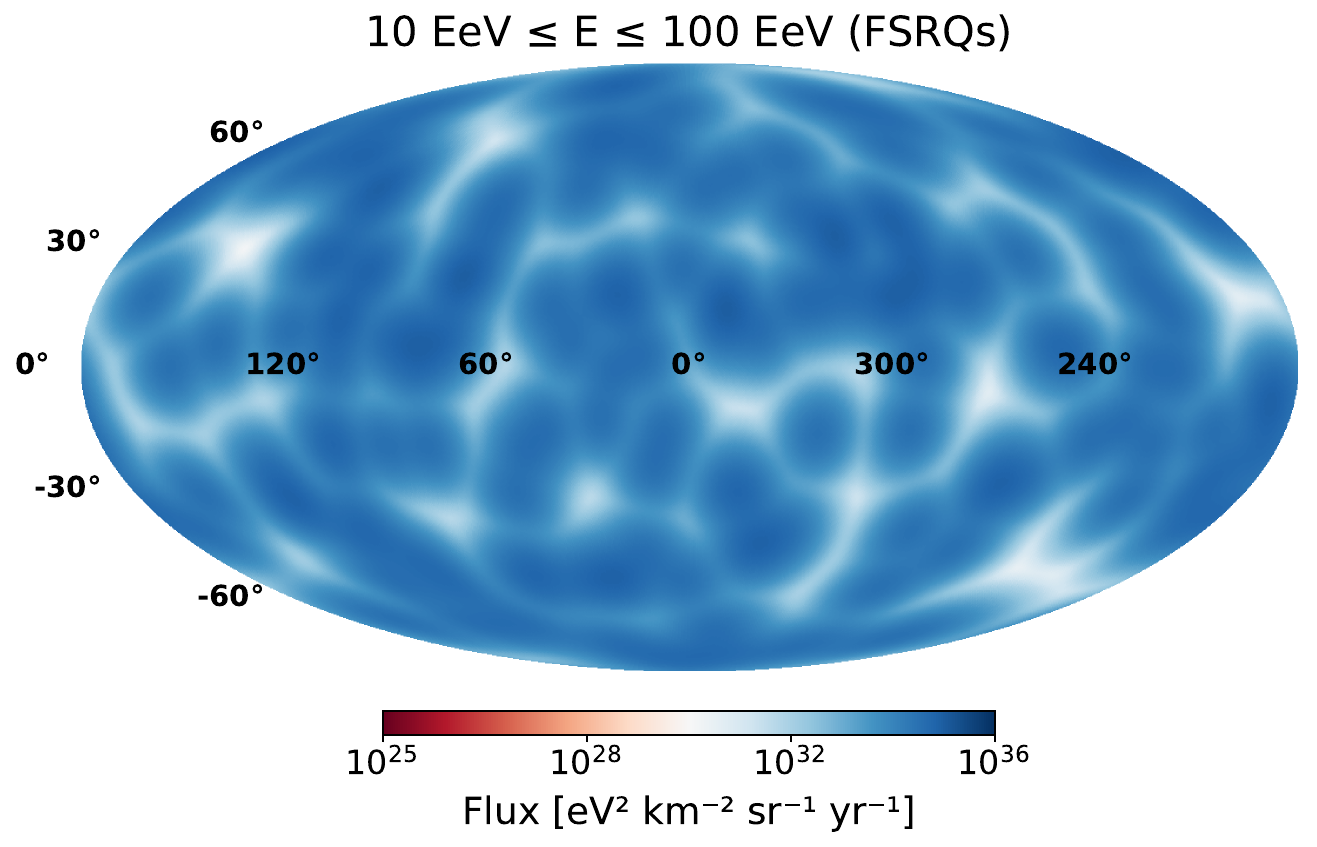}\hspace{5pt}
\includegraphics[scale=0.35]{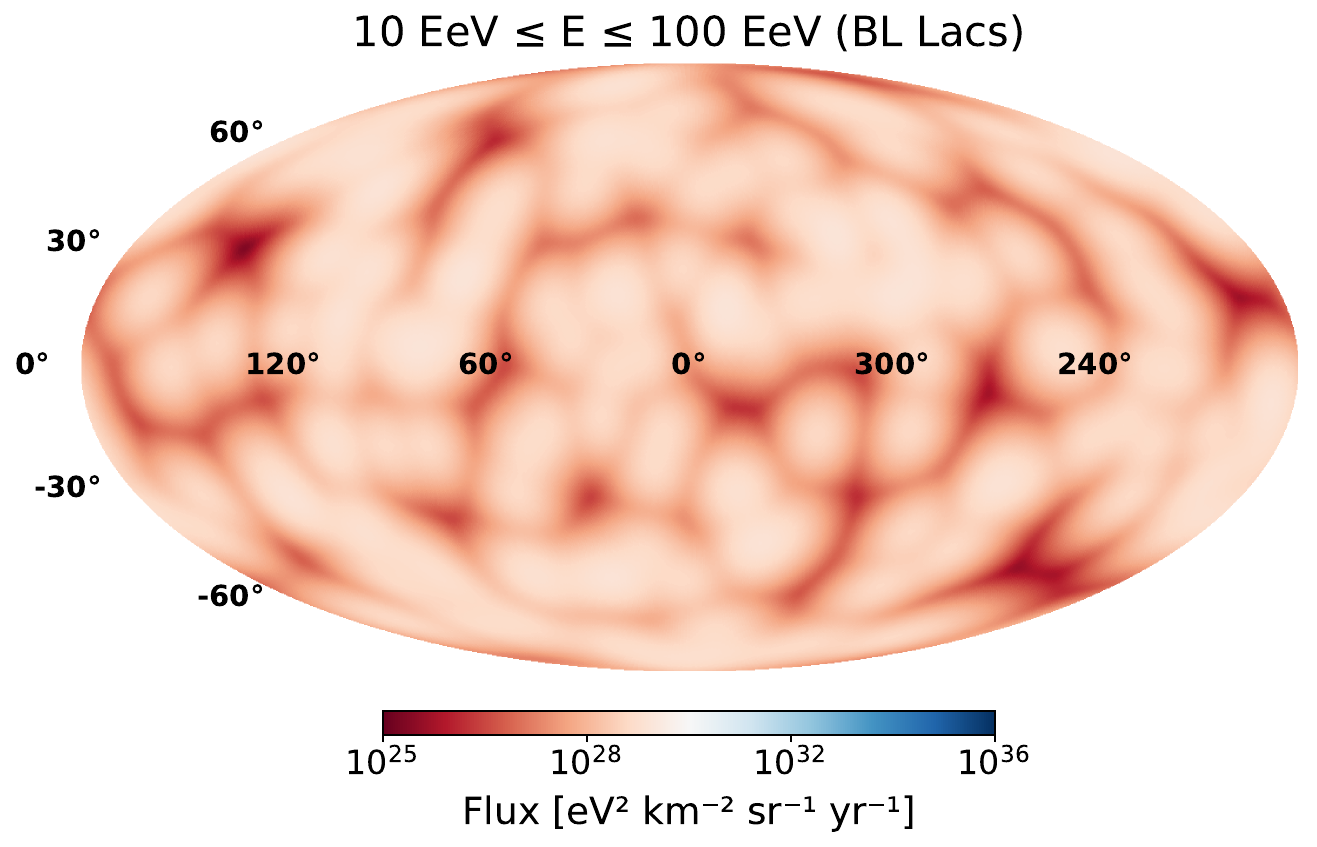}}
\caption{Mollweide skymaps of the integral UHECRs flux in the range of 
$10 \leq E \leq100~\mathrm{EeV}$ for synthetic FSRQ (left) and BL Lac (right) 
populations. The maps assume a uniform comoving distribution of hypothetical 
sources, with no observational catalogues used. A $2^\circ$ Gaussian smoothing 
is applied. The colour scale shows $E^{3}J(E)$ in 
$\mathrm{eV^{2}\, m^{-2}\, s^{-1}\, sr^{-1}}$.
}
\label{fig:uhecr_skymap}
\end{figure*}
For scenarios involving a mixed composition of CRs, the flux-weighted mean 
depth of the shower maximum is computed as
\begin{equation}
\langle X_\text{max} \rangle = \frac{\sum\limits_i J_i(E) \cdot X_\text{max, i} (E,A_i)}{\sum\limits_i J_i(E)}.
\end{equation}
Here, $J_i(E)$ denotes the flux of each nuclear species. 

We provide the $\chi^2$ test for both source models with Auger and TA data,
and it is defined as
\begin{equation}
    \chi^2 = \sum_{i} \frac{(\text{J}^i_\text{th}-\text{J}^i_\text{obs})^2}{\sigma_i^2},
\end{equation}
where $\text{J}^i_\text{th}$ are the theoretical values of flux obtained
from our calculations and $\text{J}^i_\text{obs}$ are the observed flux
values, which are obtained from the Auger and the TA experiment. $\sigma$
denotes the total uncertainty in each energy bin, incorporating both
statistical and systematic errors. The corresponding reduced $\chi^2$ values
with the combined Auger and TA data are $2.37$ and $2.55$ for FSRQs and BL Lac
objects respectively.
To assess the relative performance of different source classes, we compute the Akaike Information Criterion (AIC) and Bayesian Information Criterion (BIC), defined as
\begin{equation}
\mathrm{AIC} = \chi^2 + 2k, \quad
\mathrm{BIC} = \chi^2 + k \ln N,
\end{equation}
where $k$ is the number of free parameters and $N$ is the number of data points. These criteria provide a quantitative basis for comparing models with different parametrisations. Lower values of AIC and BIC indicate a statistically preferred model, as they represent a better balance between goodness of fit and model complexity, while higher values imply a poorer fit or unnecessary parametrisation.

Fig.~\ref{mixed1} also displays the fluxes for a mixed composition of nuclei 
along with the corresponding $\langle X_\text{max} \rangle$s and 
$\sigma \left( X_\text{max} \right )$s at the redshift $z_\text{max}=1$. The 
parametrisations show deviations from a purely proton composition, and the 
influence of the astrophysical source model is also apparent. The 
observational data for $\langle X_\text{max} \rangle$s from Auger and TA are 
taken from Refs.~\cite{xmax_pao, xmax_ta}.
These datasets provide 
$\langle X_\text{max} \rangle$ values along with statistical and 
systematic errors. The red and blue bands in 
$\langle X_\text{max} \rangle$s and $\sigma \left( X_\text{max} \right )$s 
plots denote the systematic errors of Auger and TA data, respectively. In 
the $\langle X_\text{max} \rangle$ and 
$\sigma \left( X_\text{max} \right )$ plots, the statistical uncertainties 
are shown as shaded bands. The good agreement between our results and the 
observational data supports the validity of the source models used in 
reproducing the CR spectra. For readability, we include 
Table~\ref{tab:fsrq_bllac_params}, which summarizes the key parameters 
used in fitting the UHECR energy spectrum with the Auger and TA observational
data for the FSRQ and BL Lac source classes. The best-fit value of the acceleration efficiency is $\eta \approx 0.1$. This is in good agreement with numerical simulations of particle acceleration in relativistic jets \citep{Wang:2024ijr} and magnetic reconnection \citep{Cerutti:2013mma}, which also find efficiencies of the same order. This suggests that the assumed acceleration mechanism is realistic and capable of producing the required high energies.
The parameter $d_s$ which represents an effective source separation is 
related to the source density ($d_s \equiv n_s^{-1/3}$). We obtain $d_s \simeq 160 \pm 21~\mathrm{Mpc}$ for BL Lacs, which is consistent with observational estimates from blazar luminosity functions \citep{Ajello:2011zi, Ajello:2013lka}. In contrast, the FSRQ scenario gives a much smaller value, $d_s \simeq 8 \pm 1~\mathrm{Mpc}$, implying an unrealistically high source density compared to observations \citep{Ajello:2013lka}. This indicates that FSRQs are unlikely to explain the observed UHECR flux within realistic astrophysical conditions. From Table~\ref{tab:fsrq_bllac_params}, the AIC (BIC) values for the FSRQ and BL Lac scenarios are 54.65 (60.54) 
and 50.48 (55.71), respectively. The lower values obtained for the BL Lac model indicate that it provides a better description of the data compared to the FSRQ scenario. The corresponding differences, 
$\Delta \mathrm{AIC} = 4.17$ and $\Delta \mathrm{BIC} = 4.83$, 
suggest moderate evidence in favour of the BL Lac interpretation.
The best-fit nuclear abundance includes both light and heavy nuclei, with a higher fraction of heavy elements than in the Sun, which is expected in the dense and enriched environments of AGN (see \cite{mollerachjcap}). Using the best-fit parameters, we obtain $\kappa \approx 5.65$, indicating that the required CRs power is only a few times larger than the gamma-ray luminosity and is therefore energetically viable. This value lies within the range expected for UHECR source models and indicates that the energy budget is sufficient to sustain the observed flux.

It should be mentioned that the absolute $\langle X_{\max}\rangle$ 
values measured by the Auger and the TA should not be compared directly.
As demonstrated in joint Auger-TA analyses, differences in detector design,
event-selection criteria, and reconstruction methods introduce 
experiment-dependent offsets in $\langle X_{\max}\rangle$ that are unrelated 
to mass composition. Consequently, composition inference is based on the 
behaviour of $\langle X_{\max}\rangle$ and $\sigma(X_{\max})$ relative to 
single-species expectations and on the energy dependence of these observables, 
rather than on a direct comparison between these two datasets.
\begin{table}[h]
\centering
\caption{Best-fit parameters for FSRQs and BL Lac objects. The column “Status” indicates whether a parameter is treated as free or fixed in the fitting procedure. Uncertainties correspond to approximate $1\sigma$ confidence intervals. $\varepsilon$ denotes the relative nuclear composition fractions.}
\begin{tabular}{l@{\hspace{12pt}}c@{\hspace{12pt}}c@{\hspace{12pt}}c}
\hline
\textbf{Parameter} & \textbf{Status} & \textbf{FSRQ} & \textbf{BL Lac} \\
\hline\\[-9pt]

$d_{\text{s}}$ (Mpc) & Free & $8.0 \pm 1$ & $160 \pm 21$ \\

$\eta$  & Free & $0.08 \pm 0.01$ & $0.1 \pm 0.02$ \\

$\varepsilon_\text{p}$ & Free & $0.50 \pm 0.09$ & $0.45 \pm 0.08$ \\
$\varepsilon_\text{He}$ & Free & $0.25 \pm 0.04$ & $0.28 \pm 0.06$ \\
$\varepsilon_\text{N}$ & Free & $0.15 \pm 0.02$ & $0.15 \pm 0.02$ \\
$\varepsilon_\text{Si}$ & Free & $0.06 \pm 0.002$ & $0.07 \pm 0.003$ \\
$\varepsilon_\text{Fe}$ & Free & $0.04 \pm 0.001$ & $0.05 \pm 0.001$ \\

\hline\\[-9pt]

$\chi^{2}_\text{red}$ & -- & $2.37$ & $2.55$ \\
d.o.f. & -- & $19$ & $16$ \\
AIC & -- & $54.65$ & $50.48$ \\
BIC & -- & $60.54$ & $55.71$ \\

\hline
\end{tabular}
\label{tab:fsrq_bllac_params}
\end{table}

\begin{figure*}
\centerline{
\includegraphics[scale=0.45]{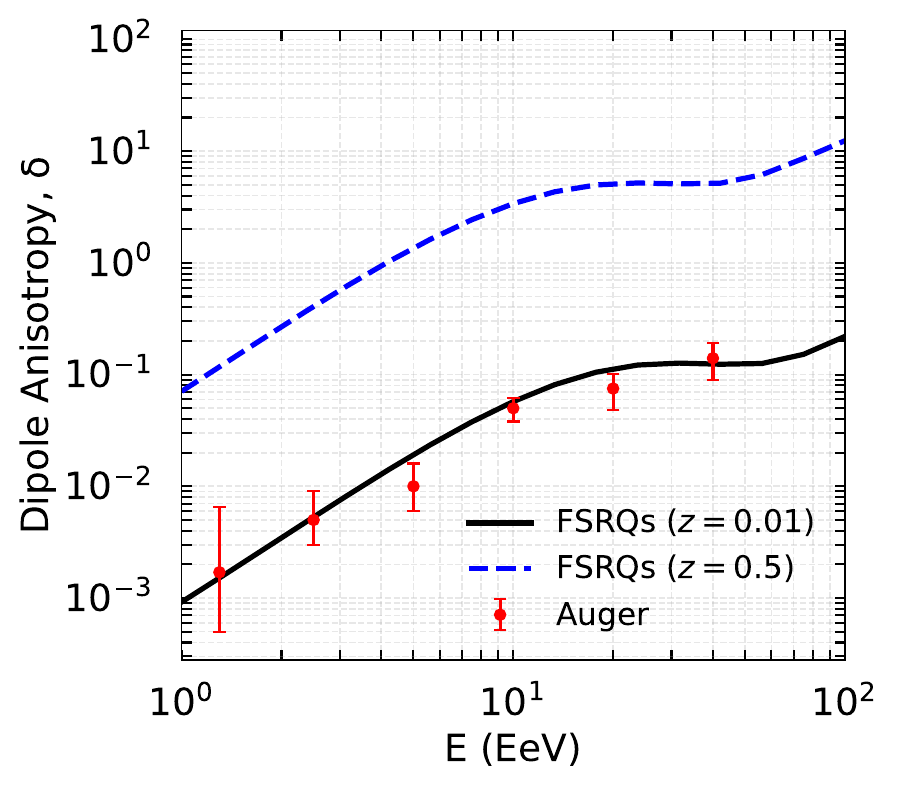}\hspace{5pt}
\includegraphics[scale=0.45]{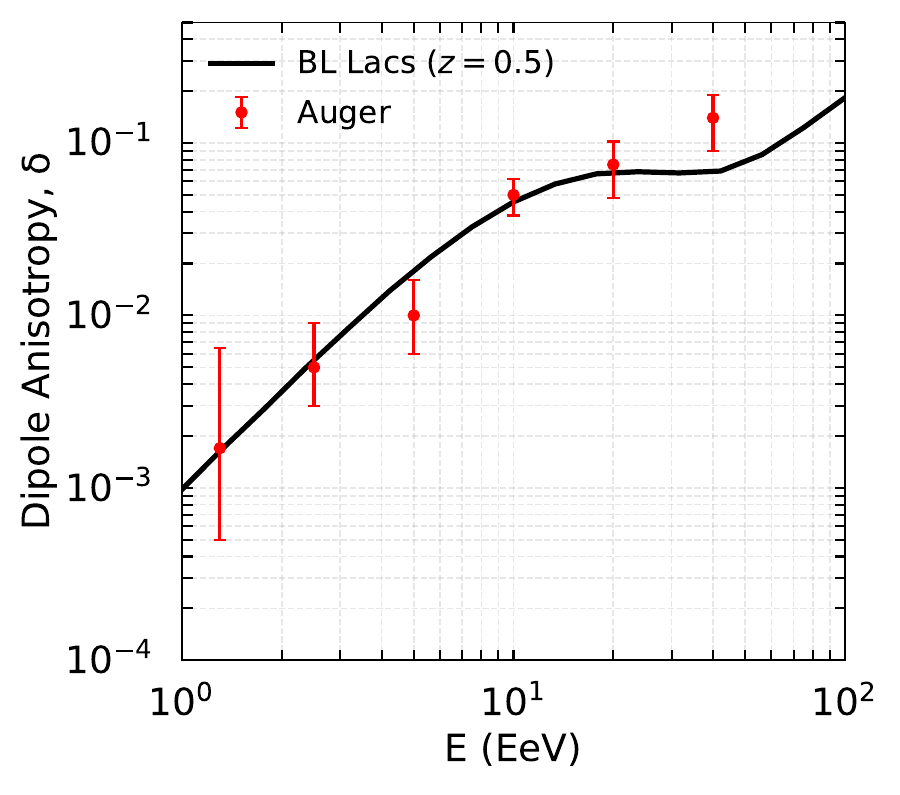}}
\caption{Dipolar anisotropy of UHECRs for FSRQs (left panel) and BL Lacs (right panel) with the observational data from Auger \citep{apj891}.
}
\label{fig:uhecr_aniso}
\end{figure*}

To describe the angular distribution of the UHECRs flux, we generate
full-sky intensity maps using the \textsc{HEALPix} framework
\citep{Gorski_2005} with a resolution parameter of ${\tt Nside}=256$,
corresponding to $N_{\text{pix}} = 12 \times N_{\text{side}}^2$ pixels
covering the full sky. In this work, all skymaps are built using a
uniform comoving distribution of synthetic sources representing the
FSRQ and BL Lac populations. No observational source catalogues are used.
The positions of the simulated sources are generated to ensure a
uniform distribution over the celestial sphere. Specifically, the
right ascension is sampled uniformly in the interval $[0,2\pi]$, while
a random variable $u$ is drawn uniformly from the range $[-1,1]$ and
the declination is obtained as $\delta = \arcsin(u)$. This procedure
guarantees an isotropic distribution of sources on the sphere. The
resulting angular coordinates are then converted into the corresponding
HEALPix coordinates $(\theta,\phi)$.
For each simulated source, we calculate its contribution to the
integral UHECR flux within the energy range
$10 \leq E \leq 100~\mathrm{EeV}$ shown in the maps. This is done by
numerically integrating the propagated energy spectrum from the chosen
energy threshold up to the maximum injection energy, and the resulting
flux is added to the corresponding HEALPix pixel.
Magnetic deflections are included through the diffusion coefficient
$D(E)$, which depends on the strength and coherence length of the
intergalactic magnetic field. These parameters in the Syrovatskii
variable $\lambda^{2}(E,z)$ describe how the flux from each source is
reduced with distance due to magnetic spreading during propagation.
To reduce pixel-to-pixel numerical noise and to make large-scale
anisotropy patterns easier to see, the final maps are smoothed using a
Gaussian function with a width of $2^\circ$.

To further test whether FSRQs and BL Lacs can explain the observed 
UHECRs, we examine the dipole anisotropy of UHECRs. The large-scale anisotropy of UHECRs is quantified by the dipole amplitude, which arises due to spatial gradients in the CR density. In the diffusive propagation regime, the dipole anisotropy is given by
\citep{harari}
\begin{equation}
\delta(E) = \frac{3D(E)}{c} \, \frac{|\nabla n(E)|}{n(E)},
\end{equation}
For a mixed composition scenario, the total dipole anisotropy is obtained as a flux-weighted sum over all species,
\begin{equation}
\delta_{\rm tot}(E) = \frac{\sum_i J_i(E)\, \delta_i(E)}{\sum_i J_i(E)},
\end{equation}
where $J_i(E)$ is the flux of the $i$-th nuclear species ($i=$ H, He, N, Si, Fe). This ensures a self-consistent treatment of both flux and anisotropy.
We plot the dipole anisotropy of UHECRs as shown in Fig.~\ref{fig:uhecr_aniso} and compare it with observational data from the Auger. The corresponding $\chi^2$ values with $z=0.5$ for FSRQs and BL Lacs are $2.69$ and $2.95$, respectively. We find that the BL Lac scenario provides a good agreement with the observed anisotropy using the same set of parameters (e.g., $z_{\max}=0.5$) that successfully reproduce the flux. In contrast, for the FSRQ case, the predicted dipole amplitude is significantly higher when using the same flux-fitted parameters ($z=0.5$). To bring the FSRQ prediction within the observed Auger limits, a much smaller redshift range ($z_{\max}=0.01$) is required. This comparison suggests that BL Lac objects provide a more consistent description of the observed UHECR flux and anisotropy, whereas FSRQs are disfavored as dominant contributors to the UHECR sky under the present framework.

\section{Summary and Conclusion}\label{summary}

\begin{figure*}
\centering
\includegraphics[scale=0.4]{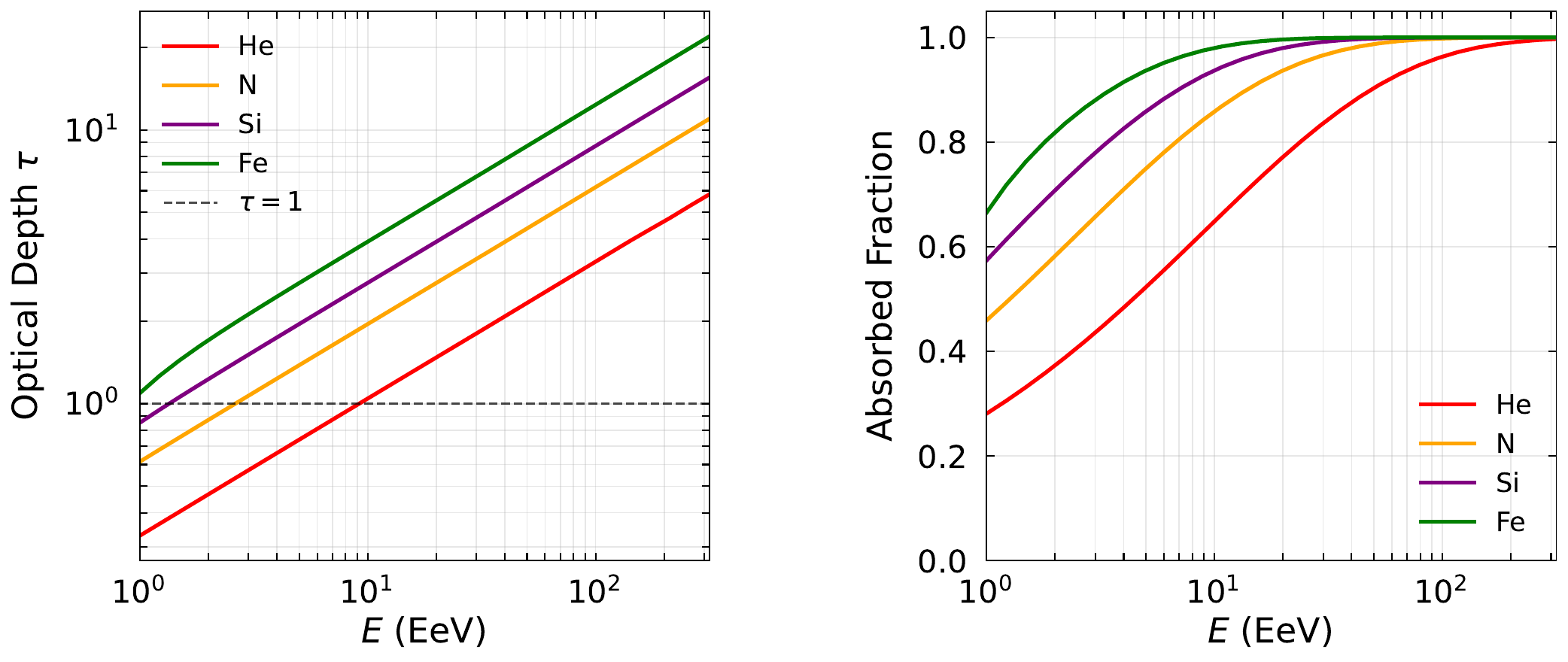}\hspace{5pt}
\caption{Photodisintegration optical depth $\tau$ (left panel) and the 
corresponding absorbed fraction (right panel) as functions of energy for 
representative primary UHECR nuclei (He, N, Si, and Fe), assuming 
propagation from sources up to redshift $z_{\max} = 0.5$ in the BL Lac 
scenario. A similar behaviour is obtained for the FSRQ scenario also (not shown here).
}
\label{fig:4}
\end{figure*}

In this work, we have investigated the flux of UHECRs within the standard 
$\Lambda$CDM cosmology, considering both FSRQs and BL Lac objects as potential 
source populations. A mixed nuclear composition consisting of p, He, N, Si, 
and Fe was adopted to model the CRs spectra and the corresponding shower 
observables $\langle X_\text{max} \rangle$ and $\sigma(X_\text{max})$. The 
theoretical predictions show a good level of agreement with the combined 
Auger and TA data, yielding reduced $\chi^2$ values of $2.37$ and $2.55$ for 
the FSRQ and BL~Lac scenarios, respectively. The deviation from a purely 
protonic composition indicates that a mixed composition provides a more 
realistic description of UHECRs propagation and air-shower behaviour. We obtain $\kappa \approx 5.65$, indicating that the required CR power is only a few times larger than the gamma-ray luminosity and is energetically consistent with UHECR source models. The best-fit acceleration efficiency $\eta \approx 0.1$ is also consistent with numerical simulations of relativistic jet acceleration and magnetic reconnection, supporting the physical plausibility of the assumed mechanism.
We find different inter-separation source distance for fitting the observable 
flux data at $8.0 \pm 1$ Mpc and $160 \pm 21$ Mpc for FSRQ and BL Lac objects respectively, 
also we get difference nuclear abundance for both sources as shown in 
Table~\ref{tab:fsrq_bllac_params}. The fitted $d_s$ value for FSRQs is not consistent
with the Fermi-LAT observation.

We constructed full-sky maps of the integral UHECR flux in the energy range
$10 \leq E \leq 100~\mathrm{EeV}$ using the \textsc{HEALPix} framework with an
angular resolution of ${\tt Nside}=256$. The maps were generated assuming a
uniform comoving distribution of synthetic sources for each source class,
without using any observational source catalogue. Source positions were
assigned randomly over the full sky, and each source contribution was added
to the corresponding \textsc{HEALPix} pixel.
The integral flux above the chosen energy threshold was
obtained by numerically integrating the propagated spectrum up to the
maximum injection energy. To suppress small-scale numerical fluctuations and
highlight large-scale anisotropy, the maps were smoothed with a Gaussian function 
with a width of $2^\circ$.

In the present work, photodisintegration of nuclei during propagation is 
treated within the continuous energy-loss (CEL) approximation through the 
total disintegration rate. The survival probability of each injected nuclear 
species is computed by integrating the corresponding photodisintegration 
interaction rate along the line of sight, including contributions from both 
the CMB and the adopted EBL model.
We note that we do not explicitly solve the full set of coupled transport 
equations describing the nuclear cascade. In particular, secondary fragments 
produced through reactions such as
$(A,Z) + \gamma \rightarrow (A-1,Z-1) + p$
or multi-nucleon emission channels are not injected as new independent 
propagating species. Instead, each primary nuclear component is propagated 
independently and attenuated according to its total interaction rate.
Because the injection model already includes light and intermediate-mass 
nuclei (H, He, N, Si, Fe), the dominant composition evolution at Earth is 
effectively captured through the differential attenuation of each species. 
In the energy range considered here ($E \gtrsim 1$ EeV), the principal 
effect of photodisintegration is the progressive suppression of heavy 
nuclei at the highest energies. This effect is properly included in our 
attenuation formalism.
A fully coupled kinetic treatment including explicit secondary production, 
as implemented in Monte Carlo propagation frameworks (e.g., CRPropa-type 
simulations), may lead to a redistribution of flux toward lighter nuclei 
at intermediate energies. Such a detailed cascade treatment will be 
explored in future work. To explicitly quantify the attenuation of 
primary nuclei during propagation within our CEL approximation, using 
Ref.~\cite{Aloisio:2013hya} we calculated the photodisintegration optical 
depth ($\tau$) and the corresponding absorbed fraction, defined as 
$1 - e^{-\tau}$. Figure~\ref{fig:4} illustrates this absorbed fraction for 
representative primary species (He, N, Si, and Fe) as a function of energy. 
The optical depth increases with both energy and nuclear mass, which means 
that heavier nuclei interact more easily during propagation. As a result, 
heavy nuclei such as iron start to undergo significant photodisintegration 
even at relatively low energies, while lighter nuclei like helium can travel 
longer distances before interacting.

We can conclude that the FSRQ scenario, characterized by strong cosmological evolution and a dominant high-redshift contribution, predicts a dipole amplitude significantly larger than observed when consistent source parameters are used. In addition, the very small source separation obtained in this case ($d_s \sim 8~\mathrm{Mpc}$) implies an unrealistically high source density compared to observational expectations, further indicating tension in the FSRQ interpretation. Attempts to reconcile these discrepancies by restricting the redshift range lead to inconsistencies with the flux, pointing to an intrinsic limitation of this scenario. In contrast, BL Lac objects, with a comparatively local distribution and weaker redshift evolution, provide a more consistent description of both the observed flux and the dipolar anisotropy. These results suggest that BL Lac objects are more favourable candidates for the origin of UHECRs, whereas FSRQs are likely subdominant contributors in this context, though they may still play an important role in multi-messenger channels such as high-energy neutrinos and gamma rays.

\section*{Acknowledgement}
SPS and UDG would like to thank the anonymous reviewer for his/her valuable feedback and comments on this work. UDG is thankful to the Inter-University Centre for Astronomy and Astrophysics 
(IUCAA), Pune, India for the Visiting Associateship of the institute.

\bibliographystyle{aasjournalv7}

\end{document}